\documentclass[aps,prl,twocolumn,showpacs,superscriptaddress,groupedaddress]{revtex4-1}
\usepackage[dvipdfmx]{graphicx}

\usepackage{amsmath}
\usepackage{stackrel,amssymb}
\usepackage[version=3]{mhchem}

\usepackage{color}

\begin{document}

\title{Optimal size for emergence of self-replicating polymer system}
\author{Yoshiya J. Matsubara}
\email{yoshi@copmlex.c.u-tokyo.ac.jp, Tel.: +81-3-5454-6732}
\author{Kunihiko Kaneko}
\email{kaneko@complex.c.u-tokyo.ac.jp, Tel.: +81-3-5454-6746}
\affiliation{Graduate School of Arts and Sciences, The university of Tokyo, 3-8-1 Komaba, Meguro-ku, Tokyo 153-8902, Japan}
\date{\today}

\begin{abstract}
A biological system consists of a variety of polymers that are synthesized from monomers, by catalysis that exists only for some long polymers. It is important to elucidate the emergence and sustenance of such autocatalytic polymerization. We analyze here the stochastic polymerization reaction dynamics, to investigate the transition time from a state with almost no catalysts to a state with sufficient catalysts. We found an optimal volume that minimizes this transition time, which agrees with the inverse of the catalyst concentration at the unstable fixed point that separates the two states, as is theoretically explained.  Relevance to the origin of life is also discussed.
\end{abstract}

\maketitle

\section{I. Introduction}

All life systems known so far consist of a wide variety of polymers that catalyze each other and are replicated through catalytic reactions. In cells, for instance, ribosomes whose main component are RNAs that synthesize a variety of protein species, such as polymerases, that catalyze RNA replication \cite{lodish2000molecular}. When considering the origins of life, it is therefore necessary to understand the emergence of a primordial polymer system that allows for self-replicating catalytic reactions, in which resource monomers such as amino acids or nucleotides, which are the building blocks of polymers, are supplied \cite{miller1953production, furukawa2015nucleobase}. It is also important to understand the timescale of the synthesis of catalytic polymers by polymerizing reactions of the monomers.
 
In this scenario, a polymer has to be long enough to function as a catalyst. In general, without catalysts, a chemical reaction to synthesize such a long polymer is extremely slow, while polymers, even if they are synthesized, are constantly degraded or diffused out. The synthesis can overcome possible degradation or diffusion only under catalysts (enzyme for protein; ribozyme for RNA) that accelerate the reaction by $10^7 \sim {10}^{19}$ \cite{wolfenden2001depth}. To sustain such a catalytically active state, a certain amount of catalysts is needed, which in turn is only synthesized from catalysts. Hence, the reaction system with autocatalytic polymers is expected to exhibit bi-stability between the inactive state with almost no catalysts and the active state with abundant catalysts that reproduce themselves. In fact, the importance of the transition from the inactive to active state for the emergence of a primitive replicating system has already been pointed out in the seminal work by Dyson \cite{dyson1985origins}, while catalytic reaction networks have also been extensively studied \cite{eigen1979hypercycle, farmer1986autocatalytic, boerlijst1991spiral, jain1998autocatalytic, segre2000compositional, kaneko2005recursive}. Here, we study this problem by considering a simple autocatalytic polymerization process, with an aim to obtain the time required for the transition from the inactive to active states.

The existence of bistable states and the transition to a catalytically active state has been discussed recently \cite{giri2012origin,wu2009origin}. In these studies, the rate equation of the concentrations of the monomers and polymers were often adopted. However, at the stage of the emergence of catalytic polymers, the number of molecules may not be so large, and the fluctuations are not negligible. Hence the description by rate equation may not be appropriate.
These fluctuations enable the transition across the barrier between the two states. Hence, we adopt a stochastic model with random collisions of molecules to investigate the transition.

The effects of fluctuations due to the smallness in molecule number have attracted considerable attention in chemical reaction dynamics \cite{togashi2001transitions,awazu2007discreteness, ohkubo2008transition, dauxois2009enhanced, biancalani2014noise, jafarpour2015noise, saito2015theoretical}. The change in the steady distribution as well as the relaxation dynamics around the steady state have been studied with respect to the decrease in the system size. However, here, we are interested in the transient time course and statistics for the transition. We study the system-size dependence of the transition time in several models for stochastic polymerization reaction dynamics, in order to find the optimal size that minimizes the transition. We then show that this time is estimated by the inverse of the concentration of the catalytic polymer at the unstable fixed point in the reaction rate equation. We will also explain the origin of this inverse law, and discuss its relevance to the origins of life.

\section{II. Autocatalytic polymer model}

To introduce our model, we postulate the following properties of the biological polymerization process. (i)Monomers are supplied sufficiently. (ii)Polymerization occurs stepwise from monomers to longer polymers. (iii)Catalytic function can emerge only for polymers with a sufficient length. (iv)Polymerization reaction proceeds extremely slowly without catalysts, but is accelerated drastically with them.

Initially only monomers are supplied, and the polymerization progresses slowly owing to degradation or diffusion. Once sufficient catalysts are synthesized, the catalytic reaction to synthesize them progress constantly.  We investigate the emergence and time scale of such autocatalysis by introducing a model consisting of polymers. We denote the polymer of length $n$ as $A(n)$, with $A(1)$ being a supplied monomer. Sequence information under multiple monomer species is disregarded, and only the length is considered. We assume that catalytic capacity appears at $n=L$. For successive polymerization for each ligation reaction, we introduce model (I) with a set of reversible reactions:

\begin{eqnarray*}
A(n-1) + A(1) &\underset{1}{\overset{\alpha}{\leftrightarrows}}& A(n) \\
A(n-1) + A(1) + A(L) &\underset{k}{\overset{\alpha k}{\leftrightarrows}}& A(n) + A(L) \\
A(n) &\overset{\mu}{\rightarrow}& \emptyset \qquad (n = 2,3,4, \dots, L)
\end{eqnarray*}

Conversely, for polymerization processes which double the length, we introduce model (II):

\begin{eqnarray*}
A(n) + A(n) &\underset{1}{\overset{\alpha}{\leftrightarrows}}& A(2n) \\
A(n) + A(n) + A(L) &\underset{k}{\overset{\alpha k}{\leftrightarrows}}& A(2n) + A(L) \\
A(n) &\overset{\mu}{\rightarrow}& \emptyset \qquad (n = 2,4,8, \dots, L)
\end{eqnarray*}

We first consider the simplest case for model (II), i.e., $L=4$; the general case will be studied later.

The model consists only of monomers, dimers, and tetramers. Two monomers ligate into a dimer, and dimers ligate into a tetramer. Without losing generality, the spontaneous reaction rate without catalysts is set at unity, while the ligation reaction is accelerated $k$ times by the catalytic tetramers, where $k \gg 1$, accordingly. The backward reaction rate is set to $\alpha$. We present the case where $\alpha=1$, while for $\alpha \geq 1$ the result generality holds. Indeed, the synthesis of the catalytic polymer often needs energetic cost, and the forward reaction is slower than the backward one, and then $\alpha \geq 1$ is resulted. Even if $\alpha<1$, however, as long as the polymer concentration decreases with its length, the discussed results are valid.

These molecules are placed in a container with a (constant) volume. We assume that molecules are well mixed in the container, and we do not consider spatial inhomogeneity. From the external reservoir, monomers are supplied sufficiently fast, such that its number $n_1 (=V)$ is assumed to be constant. Then, our dynamics are represented by the number of dimers $n_2(t)$ and tetramers $n_4(t)$ at time $t$. The dimers and tetramers are diffused out so that their numbers decrease with the rate $\mu$. Considering that all the reaction processes occur stochastically with the rate following the mass action law for reaction, the dynamics of the probability distribution of $n_2$ and $n_4$ is given by

\begin{equation} \label{eq:master}
\begin{aligned}
\partial P(n_2, n_4, t) /\partial t = \\ 
  \sum_{m=1}^6 \Bigl[ \pi_m  P(n_2-\nu&_{2,m}, n_4 -\nu_{4,m}, t) - \pi_m  P(n_2, n_4, t) \Bigr],
\end{aligned}
\end{equation}

where $m$ is the index of the reaction, and $\nu_{2,m},\nu_{4,m}$ are the increment in the number of dimers and tetramers in the $m$th reaction (for example, $\nu_{2,1}=1,\nu_{4,2}=0,\nu_{2,2}=-1,\nu_{4,2}=0$, and so forth). The first term describes the inflow of the probability from $(n_2 -\nu_{2,m}, n_4 -\nu_{4,m} )$ by the reaction, and the second term describes the outflow of probability from $(n_2, n_4)$. Each $\pi_m (\nu_{2,m},\nu_{4,m})$ is the transition probability of monomers ligation, dimer cleavage, dimers ligation, tetramer cleavage, diffusion of dimer and tetramer, respectively, as given by

\begin{equation*}
\begin{aligned}
&\pi_1(+1,0)= \frac{1}{2} \kappa n_1 ( n_1 -1) V^{-1}, \pi_2(-1,0)=\alpha \kappa  n_2 \\
&\pi_3(-2,+1)=\frac{1}{2} \kappa n_2 (n_2-1) V^{-1} \\
&\pi_4(+2,-1)= \alpha n_4 V^{-1} + \alpha k n_4 (n_4-1) V^{-1} \\
&\pi_5(-1,0)=\mu n_2, \pi_6(0,-1)=\mu n_4,
\end{aligned}
\end{equation*}

where $\kappa$ is the sum of the rate of spontaneous and catalytic reaction as: $\kappa = 1 + k n_4 V^{-1}.$

In the continuum limit with $V, n_i\rightarrow \infty$, the change in molecule concentration ($x_1 = 1, x_2 = n_2/V, x_4 = n_4/V$) \cite{about-unit} is represented by the deterministic rate equation \cite{van1992stochastic}, as given by

\begin{equation} \label{eq:rate_eq}
\begin{aligned}
\dot{x}_2 =& (1+k x_4) ( \frac{1}{2} x_1^2- \alpha x_2-x_2^2+ \alpha 2 x_4) &- \mu x_2 \\
\dot{x}_4 =& (1+k x_4) ( \frac{1}{2} x_2^2- \alpha x_4) &- \mu x_4
\end{aligned}
\end{equation}

\begin{figure}
\centering
\includegraphics[width=5cm]{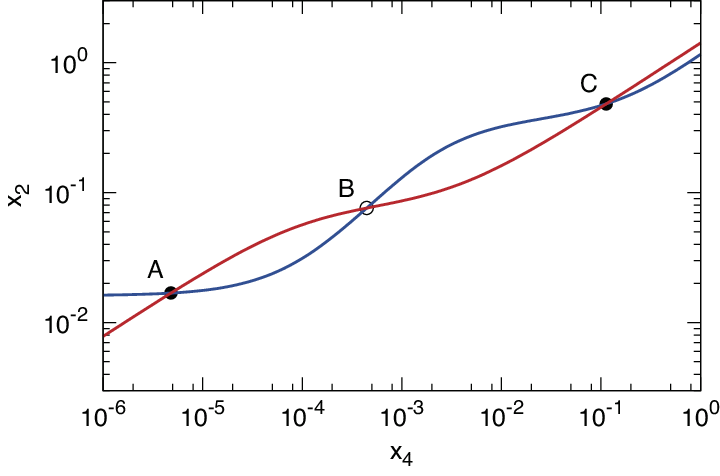}
\caption{
Nullclines of the rate equation Eq.\ref{eq:rate_eq} for  $k=10^4$ and $\mu=30$, plotted in the phase space $(x_4,x_2)$. The blue line represents the nullcline $\dot{x_2}=0$, and the red line represents the nullcline $\dot{x_4}=0$. Intersection points of the lines $A$, $B$, and $C$ are fixed points of the equation, where $A$ and $C$ (black circle) are stable fixed points and $B$ (white circle) is an unstable one.
}
\label{fig:nullclines}
\end{figure}

For a certain range of parameters $k$ and $\mu$, this deterministic rate equation has bi-stable fixed points, separated by an unstable fixed point, which are denoted by $A(x_2^a, x_4^a), B(x_2^b, x_4^b)$ , and $ C(x_2^c, x_4^c), (x_2^a<x_2^b<x_2^c, x_4^a<x_4^b<x_4^c)$ (see Fig.\ref{fig:nullclines}). The fixed point $A$ corresponds to the non-active state with only few dimers and tetramers, while $C$ corresponds to the active state with autocatalytic reactions and sufficient catalysts.

\section{III. Optimal volume for the transition and its relation with unstable state}

\begin{figure}
\centering
\includegraphics[width=8cm]{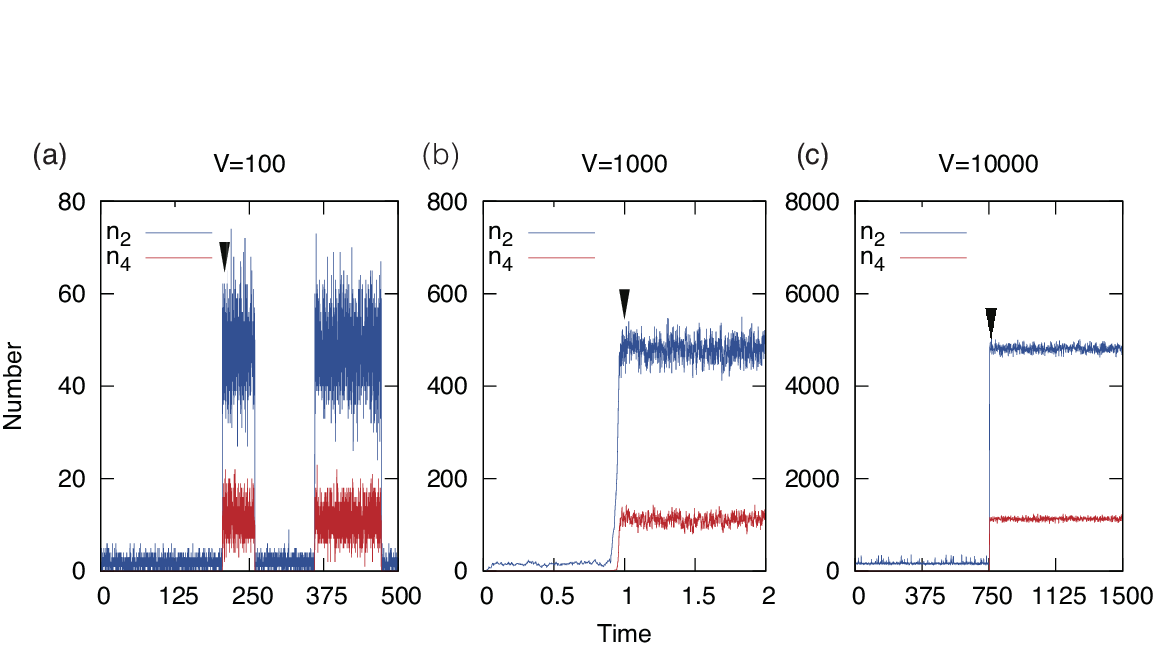}
\caption{
Time series of the number of dimers $n_2$ (blue line) and tetramers $n_4$ (red line) for $k=10^4,\mu=30$ and (a)$V=10^2$,(b)$10^3$,(c)$10^4$. The time for the transition to the active state are $t^* \sim 204$, $\sim 1$, and $\sim 754$  for (a),(b), and (c), respectively.
}
\label{figtimeseries}
\end{figure}
\begin{figure}
\centering
\includegraphics[width=5cm]{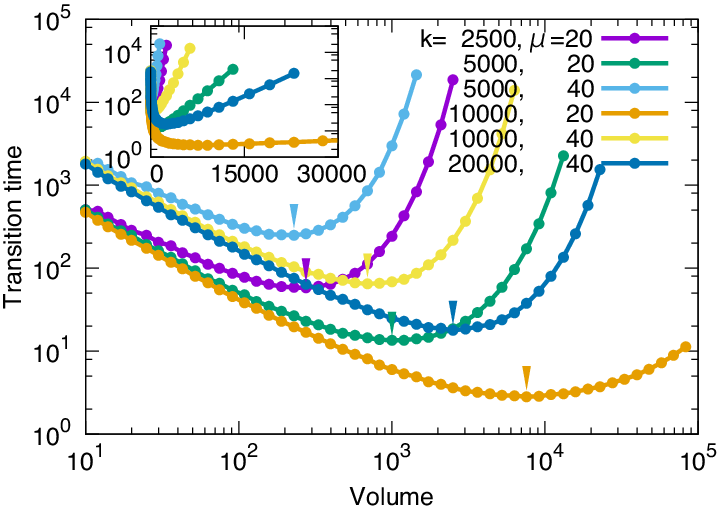}
\caption{
The transition time from the inactive to active states, plotted as a function of the volume $V$. For each parameter and size, the time is computed as the average over $10^4$ samples of the master equation. Log--log plot for $(k,\mu) =(2.5\times10^3,20), (5\times10^3,20), (5\times10^3,40),  (10^4,20), (10^4,40) $and$ (2\times10^4,40)$, with different colors and symbols. The optimal $V_{opt}$ that gives the minimum is indicated by the thick arrow. The inset shows the corresponding semi-log plot.
}
\label{fig:vdependb}
\end{figure}
\begin{figure}
\centering
\includegraphics[width=4cm]{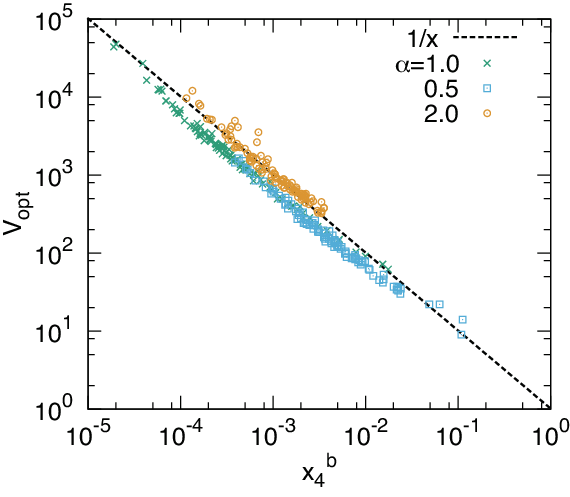}
\caption{
The relationship between $V_{opt}$ and $x_4^b$. The abscissa give the tetramer concentration at the unstable fixed point $x_4^b$, and the ordinate shows the optimal volume $V_{opt}$ as estimated in Fig.\ref{fig:vdependb}. Each point is a result from different parameter values randomly selected from $k \in (0,5\times{10}^4), \mu \in (0,100)$ where the rate equation exhibits bi-stability. }
\label{fig:xbvopt}
\end{figure}

We now study the probabilistic chemical reaction system by the Gillespie algorithm \cite{gillespie1977exact}. Examples of the time series of the number of molecules are shown in Fig.\ref{figtimeseries}, for different values of $V$, where the initial condition of the number of dimers and tetramers is set to zero. As shown, the transition from the inactive to active state is observed at a certain time, while for small $V$, there also exists transition back from the active to inactive state.

We now define the transition time to the active state by the time when the concentrations of both dimers and tetramers $n_2/V, n_4/V$  exceed the value given by the active fixed point $C(x_2^c,x_4^c)$ in the rate equation. The volume dependency of the transition time is plotted in Fig.\ref{fig:vdependb}. As can be seen in the figure there is a minimum transition time at a certain optimal volume $V_{opt}$. By decreasing $V$ below $V_{opt}$, the transition time increases asymptotically as $V^{-1}$, while for $V> V_{opt}$, it increases exponentially with the volume.

We computed $V_{opt}$ by varying the parameters $k$ and $\mu$. Interestingly, as shown in Fig.\ref{fig:xbvopt}, these optimal volumes, when plotted as a function of $x_4^b$, i.e., the tetramer concentration at the unstable fixed point in rate equation, are fitted on a single curve, given by $V_{opt} \sim 1/x_4^b$ \cite{about-unitconversion}, for all the parameter values. (If backward reaction rate $\alpha$ is close to zero, the polymer concentration could be higher than the monomer at the unstable fixed point, and in this case, the monomer number that satisfies $1/x_4^b$ is less than unity for small $V$, and the optimal volume cannot exist. Apart from this unrealistic case this relationship of the optimal volume generally holds.)

We now explain the origin of this optimal volume and its relationship with $x_4^b$. First of all, by using the standard formalism \cite{van1992stochastic} master equation is approximated by the Fokker--Planck equation (FP) for large volume. It is obtained by the Kramers--Moyal expansion or the chemical Langevin equation, as the expansion by $1/V$.  For our model ($\alpha = 1, x_1 = 1$), the FP equation is straightforwardly obtained as

\begin{equation} \label{eq:fokker_planck}
\begin{aligned}
\partial &P(x_2, x_4, t) / \partial t = - \sum_{i=2,4} \partial_{x_i} (A_i(x) P(x, t) )  \\
&\qquad \qquad \qquad+ \frac{1}{2V} \sum_{i,j=2,4} \partial_{x_i} \partial_{x_j} (B_{i,j}(x) P(x,t) ) \\
&A = \begin{pmatrix} (1 + k x_4 )( \frac{1}{2}  - x_2 - x_2^2 + 2 x_4) - \mu x_2 \\
 (1 + k x_4 )( \frac{1}{2} x_2^2 - x_4) - \mu x_4
\end{pmatrix} \\
&B = \begin{pmatrix} (1 + k x_4 )( \frac{1}{2}  + x_2) + \mu x_2 &
4 (1+k x_4) ( \frac{1}{2} x_2^2 + x_4)\\
 (1 + k x_4 )( \frac{1}{2} x_2^2 + x_4) & \mu x_4
\end{pmatrix}
\end{aligned}
\end{equation}

\begin{figure}
\centering
\includegraphics[width=5cm]{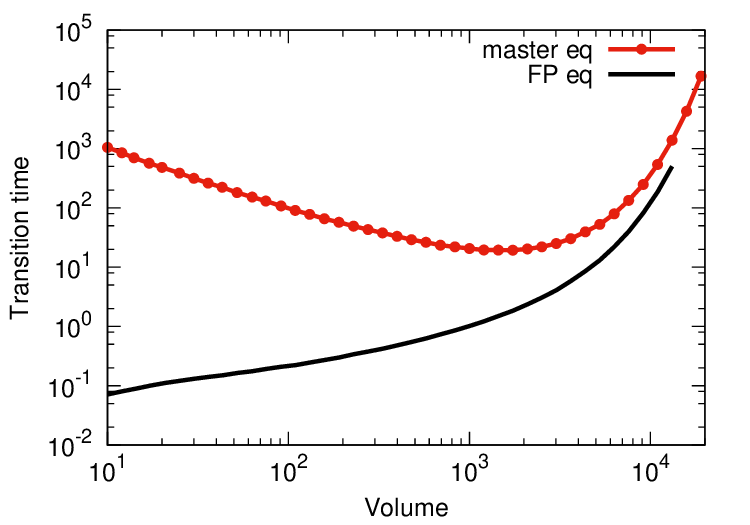}
\caption{The transition time estimated by the FP equation Eq.\ref{eq:fokker_planck} for $k=10^4$,$\mu=30$ (black line); the transition time directly estimated by the master equation is also shown for comparison (red line).}
\label{fig:fokker}
\end{figure}

We computed the volume dependency of the transition time from the initial value $x_2=x_4=0$ to $x_2 > x_2^c, x_4 > x_4^c$ by using the FP equation, as is plotted in  Fig.\ref{fig:fokker}. The transition time increases monotonically with $V$, and is fitted well with $\exp(const.\times V)$. This is expected from the Kramers formula \cite{gardiner1985handbook} in which the probability to jump from one stationary state to another is given by $\exp(- \epsilon U)$, where $\epsilon$ is the noise strength and $U$ is the potential barrier to go to the new stationary state. Here the noise in the FP equation is proportional to $\epsilon=1/V$, and the transition to the active state which has to go across the unstable fixed point is given by the jump over the potential barrier $U_b$. Hence the transition time increases as $\exp(U_b V)$ with the increase in $V$. In fact, the transition time computed by the master equation asymptotically agrees with this estimate for large $V$.

(Indeed, for a one-variable Fokker--Planck equation
\begin{equation}
 \partial_t P(x,t) = - \partial_x \bigl( A(x) P(x,t)  \bigr) + \frac{1}{2} \partial_x^2 \bigl( B(x) P (x,t) \bigr),
 \end{equation}
 the corresponding
  potential is given by $U(x)=\int^x A(x')/B(x') dx'$. In the present two-variable case, such potential does not generally exist \cite{graham1984weak}, but the temporal evolution to cross the saddle-point $B$ is restricted around its unstable manifold that is located between the two nullclines (see Fig.\ref{fig:nullclines}). Thus the motion is effectively represented by a one-dimensional motion, and the effective potential exists for the present Fokker--Planck equation.)
 
As this dependence is monotonic in $V$, the FP equation can explain neither the existence of $V_{opt}$ nor the increase in the transition time for $V<V_{opt}$. For the explanation the discreteness of the molecule number $0,1,2,\dots$ is very important.  We now explain the relationship $V_{opt} \sim 1/x_4^b$ along this line.

For the transition to occur, the concentration $x_4$ has to exceed the value of the unstable fixed point $x_4^b$, while the catalytic reaction needs at least a single tetramer. For small $V<1/x_4^b$, the average tetramer number is decreased below unity. The tetramer number is therefore often zero, and the probability of having a single molecule decreases with the decrease in $V$. Then the probability of crossing $x_4^b$ decreases with $V$, if $Vx_4^b<1$. Thus, the transition time increases as $V$ decreases below $1/x_4^b$, which gives the optimal volume for the transition time. 

We have confirmed that the relationship between the optimal volume and the unstable fixed point is universal in an autocatalytic polymerization process, as given by models (I) and (II). Indeed, most autocatalytic polymerization process can be essentially constructed by combination of models (I) or (II). Here, the master equations for probabilities are given in the same way as in Eq.\ref{eq:master}, and the rate equation in the infinite volume limit for models (I) and (II) is given by:

\begin{equation} \label{eq:modelA}
\begin{aligned}
\dot{x}_n &= (1+k x_{L}) ( x_{n-1} x_1 -x_n -x_n  x_1+ x_{n+1}) - \mu x_n \\ &\qquad \qquad \qquad \qquad(n = 2,3,4, \dots, L-1)\\
\dot{x}_{L} &= (1+k x_{L}) ( x_{{L-1}} x_1 -x_{L}) - \mu x_{L},
\end{aligned}
\end{equation} for model (I) and
\begin{equation} \label{eq:modelB}
\begin{aligned}
\dot{x}_n &= (1+k x_{L}) ( \frac{1}{2}x_{n/2}^2-x_n-x_n^2+2 x_{2n}) - \mu x_n \\ &\qquad \qquad \qquad \qquad(n = 2,4,8, \dots, L/2)\\
\dot{x}_{L} &= (1+k x_{L}) ( \frac{1}{2} x_{{L/2}}^2-x_{L}) - \mu x_{L}.
\end{aligned}
\end{equation}
for model (II).

\begin{figure}
\centering
\includegraphics[width=5cm]{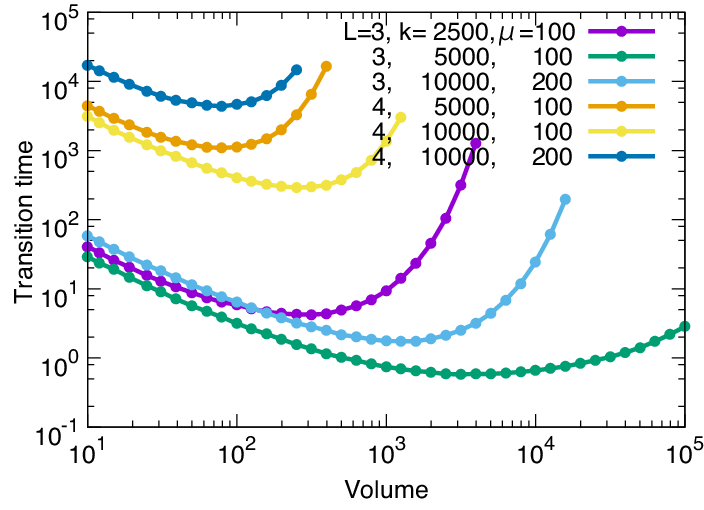}
\caption{
The same plot for model (I) with $(L,k,\mu)=(3,2.5\times 10^3, 100),(3,5.0\times 10^3, 100),(3,1.0\times 10^4, 200), (4,5.0\times 10^3, 100),(4,1.0\times 10^4, 100),(4,1.0\times 10^4, 200)$. Computed in the same manner as Fig.\ref{fig:vdependb}
}
\label{fig:vdependa}
\end{figure}
\begin{figure}
\centering
\includegraphics[width=4cm]{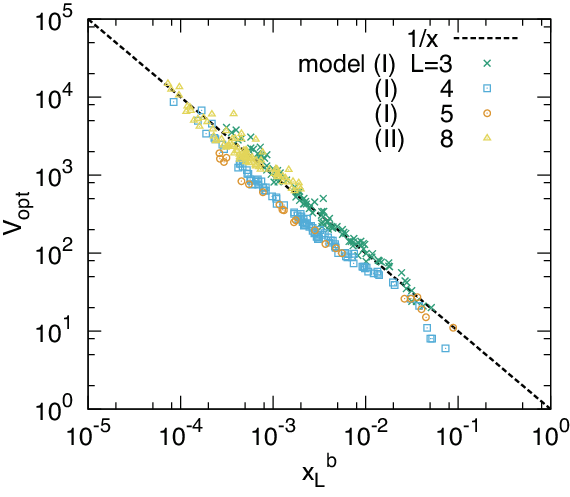}
\caption{
The relationship between $V_{opt}$ and $x_L^{b}$ for model (I) with $L=3$ (green), $4$ (blue) and $5$ (orange) and for model (II) with $L=8$ (yellow). The data are computed in the same way as for Fig.\ref{fig:xbvopt}. Parameter values are randomly selected from $k \in (0, {10}^4), \mu \in (0,100)$. }
\label{fig:xbvoptgen}
\end{figure}

Again, the dynamical systems of the model equations (\ref{eq:modelA}) and (\ref{eq:modelB}) have two stable fixed points $ A(x_i^a), C(x_i^c)$ and one unstable fixed point$ B(x_i^b)$, for a certain parameter region. Fixed points in model (II) are the root of the self-consistent equation: $$ x_L = \left( \frac{1 + k x_L}{\alpha(1 + k x_L) + \mu} \right)^{L-1}. $$ Using direct Gillespie simulations of the model, we computed the transition time between the state with $ x_{L}  \sim 0$ to the active state with $ x_{L}  \sim x_L^c$, which has a minimum at a certain volume $V_{opt}$ (see Fig.\ref{fig:vdependa}). 
We have plotted $V_{opt} $ for a variety of parameters and maximal lengths $L$, again as a function of $x_L^{b}$ in Fig.\ref{fig:xbvoptgen}. We have thus confirmed that the relationship between the optimal volume and the catalyst concentration at the unstable fixed point, $V_{opt} \sim 1/x_L^{b}$ is universal.

Last, we have also studied a model with plural monomer species and sequence dependent catalytic reaction. In this case again, we have confirmed the existence of optimal size for the transition to the active catalytic state, where specific sequences are selected.

\section{IV. Summary and Discussion}

To sum up, we have unveiled the explicit condition for autocatalytic polymers to emerge. As spontaneous synthesis of polymers is much slower than the diffusion or degradation, longer polymers are not maintained only by that. In contrast to this inactive state, catalysts are continuously replicated with the autocatalytic reaction, in the active sate. Fluctuations in the number of molecules induce the transition from the former to the latter state while at least a single catalyst is needed for it. This tradeoff leads to the optimal volume to minimize the transition time, generally given by the inverse of the catalyst concentration at the unstable fixed point of the deterministic rate equation. Below this optimal volume, the time to assemble a single catalytic polymer increases in proportion to $V^{-1}$, while beyond it, more catalytic polymers are needed for the transition, and the time to achieve such fluctuation increases exponentially with $V$. 

Discreteness (0,1,..) in the molecule number essentially matters for the determination of the optimal volume, in contrast to several studies that adopt the FP equation by the system-size expansion\cite{van1992stochastic,gillespie2000chemical} to obtain the change in distribution\cite{biancalani2014noise,saito2015theoretical} and the transition process \cite{duncan2015noise, haruna2010investigating}. In fact, the FP equation approach can account for the exponential increase in the transition time with the system size but not the optimal volume. It is also interesting to note that despite the importance in the discreteness in the number, the optimal volume is estimated by the unstable fixed point of the {\em rate equation} that itself is obtained in the infinite size limit.

Our result applies generally to an autocatalytic replication process, where inactive and active states are bistable.  The inverse relationship between the optimal volume and the catalyst concentration at the unstable fixed point that separates the two stable states is general even though the unstable-point concentration depends on several reaction parameters, and the details of the polymerization process. In models (I) and (II), the concentration decreases exponentially with the (minimum) length  of the catalyst polymer, such that the optimal volume as well as the transition time increases.

Although the emergence of life will require several steps, autocatalytic polymerization is essential as one of them. The present result suggests that the volume of the reaction space matters for it. A limited space with an appropriate size is preferable, which would be first provided by a porous medium such as a mineral surface, and later by a vesicle composed by the lipid bilayer. For example, by considering the model (II) with $k={10}^{8} \mathrm{[L/mol]}$, $\mu = 1$ (with the unit of spontaneous reaction rate), $\alpha = 1 \mathrm{[L/mol]}$, $x_1 = 1 \mathrm{[mol/L]}$ and $L = 16$ the unstable fixed point value of the catalyst in the rate equation becomes $x_{L}^b = 8.4 \times 10^{-9} \mathrm{[mol/L]}$. Therefore, $V_{opt} \sim 1/ (x_{L}^b N_A )$ is estimated by $2 \times { 10}^{-16} \mathrm{ [L]} $, which is as small value as a bacterial cell,  while these parameter values are not unique and reliable estimates are difficult since the chemical parameters for the primitive catalytic polymerization are not currently available. This estimate, however, will be useful to design protocells with catalytic polymers within a vesicle, which has been extremely investigated in synthetic biology \cite{luisi2006emergence,rasmussen2009protocells}. Our optimal size will provide a guide to choose an appropriate vesicle size.

\section{ acknowledgement}

We thank Nen Saito, Yuki Sughiyama, Tetsuhiro S. Hatakeyama, Nobuto Takeuchi and  Atsushi Kamimura for the fruitful discussions. This research is partially supported by the Platform Project for Supporting in Drug Discovery and Life Science Research (Platform for Dynamic Approaches to Living System) from Japan Agency for Medical Research and Development(AMED).


\begin{thebibliography}{33}%
\makeatletter
\providecommand \@ifxundefined [1]{%
 \@ifx{#1\undefined}
}%
\providecommand \@ifnum [1]{%
 \ifnum #1\expandafter \@firstoftwo
 \else \expandafter \@secondoftwo
 \fi
}%
\providecommand \@ifx [1]{%
 \ifx #1\expandafter \@firstoftwo
 \else \expandafter \@secondoftwo
 \fi
}%
\providecommand \natexlab [1]{#1}%
\providecommand \enquote  [1]{``#1''}%
\providecommand \bibnamefont  [1]{#1}%
\providecommand \bibfnamefont [1]{#1}%
\providecommand \citenamefont [1]{#1}%
\providecommand \href@noop [0]{\@secondoftwo}%
\providecommand \href [0]{\begingroup \@sanitize@url \@href}%
\providecommand \@href[1]{\@@startlink{#1}\@@href}%
\providecommand \@@href[1]{\endgroup#1\@@endlink}%
\providecommand \@sanitize@url [0]{\catcode `\\12\catcode `\$12\catcode
  `\&12\catcode `\#12\catcode `\^12\catcode `\_12\catcode `\%12\relax}%
\providecommand \@@startlink[1]{}%
\providecommand \@@endlink[0]{}%
\providecommand \url  [0]{\begingroup\@sanitize@url \@url }%
\providecommand \@url [1]{\endgroup\@href {#1}{\urlprefix }}%
\providecommand \urlprefix  [0]{URL }%
\providecommand \Eprint [0]{\href }%
\providecommand \doibase [0]{http://dx.doi.org/}%
\providecommand \selectlanguage [0]{\@gobble}%
\providecommand \bibinfo  [0]{\@secondoftwo}%
\providecommand \bibfield  [0]{\@secondoftwo}%
\providecommand \translation [1]{[#1]}%
\providecommand \BibitemOpen [0]{}%
\providecommand \bibitemStop [0]{}%
\providecommand \bibitemNoStop [0]{.\EOS\space}%
\providecommand \EOS [0]{\spacefactor3000\relax}%
\providecommand \BibitemShut  [1]{\csname bibitem#1\endcsname}%
\let\auto@bib@innerbib\@empty
\bibitem [{\citenamefont {Lodish}\ \emph {et~al.}(2000)\citenamefont {Lodish},
  \citenamefont {Berk}, \citenamefont {Zipursky}, \citenamefont {Matsudaira},
  \citenamefont {Baltimore}, \citenamefont {Darnell} \emph
  {et~al.}}]{lodish2000molecular}%
  \BibitemOpen
  \bibfield  {author} {\bibinfo {author} {\bibfnamefont {B.}\ \bibnamefont
  {Alberts}} \emph {et~al.},\ }\href@noop {}
  {\emph {\bibinfo {title} {Molecular Biology of the Cell}}}, 6th ed.  (\bibinfo  {publisher} {Garland Science},\ \bibinfo {year} {2014})\BibitemShut
  {NoStop}%
\bibitem [{\citenamefont {Miller}\ \emph {et~al.}(1953)\citenamefont {Miller}
  \emph {et~al.}}]{miller1953production}%
  \BibitemOpen
  \bibfield  {author} {\bibinfo {author} {\bibfnamefont {S.~L.}\ \bibnamefont
  {Miller}} \emph {et~al.},\ }\href@noop {} {\bibfield  {journal} {\bibinfo
  {journal} {Science}\ }\textbf {\bibinfo {volume} {117}},\ \bibinfo {pages}
  {528} (\bibinfo {year} {1953})}\BibitemShut {NoStop}%
\bibitem [{\citenamefont {Furukawa}\ \emph {et~al.}(2015)\citenamefont
  {Furukawa}, \citenamefont {Nakazawa}, \citenamefont {Sekine}, \citenamefont
  {Kobayashi},\ and\ \citenamefont {Kakegawa}}]{furukawa2015nucleobase}%
  \BibitemOpen
  \bibfield  {author} {\bibinfo {author} {\bibfnamefont {Y.}~\bibnamefont
  {Furukawa}}, \bibinfo {author} {\bibfnamefont {H.}~\bibnamefont {Nakazawa}},
  \bibinfo {author} {\bibfnamefont {T.}~\bibnamefont {Sekine}}, \bibinfo
  {author} {\bibfnamefont {T.}~\bibnamefont {Kobayashi}}, \ and\ \bibinfo
  {author} {\bibfnamefont {T.}~\bibnamefont {Kakegawa}},\ }\href@noop {}
  {\bibfield  {journal} {\bibinfo  {journal} {Earth and Planetary Science
  Letters}\ } \textbf {\bibinfo {volume} {429}},\ \bibinfo {pages}
  { 216} (\bibinfo {year} {2015})}\BibitemShut {NoStop}%
\bibitem [{\citenamefont {Wolfenden}\ and\ \citenamefont
  {Snider}(2001)}]{wolfenden2001depth}%
  \BibitemOpen
  \bibfield  {author} {\bibinfo {author} {\bibfnamefont {R.}~\bibnamefont
  {Wolfenden}}\ and\ \bibinfo {author} {\bibfnamefont {M.~J.}\ \bibnamefont
  {Snider}},\ }\href@noop {} {\bibfield  {journal} {\bibinfo  {journal}
  {Accounts of chemical research}\ }\textbf {\bibinfo {volume} {34}},\ \bibinfo
  {pages} {938} (\bibinfo {year} {2001})}\BibitemShut {NoStop}%
\bibitem [{\citenamefont {Dyson}(1985)}]{dyson1985origins}%
  \BibitemOpen
  \bibfield  {author} {\bibinfo {author} {\bibfnamefont {F.~J.}\ \bibnamefont
  {Dyson}},\ }\href@noop {} {\emph {\bibinfo {title} {Origins of life}}}\
  (\bibinfo  {publisher} {Cambridge University Press},\ \bibinfo {year}
  {1985})\BibitemShut {NoStop}%
\bibitem [{\citenamefont {Eigen}\ and\ \citenamefont
  {Schuster}(1979)}]{eigen1979hypercycle}%
  \BibitemOpen
  \bibfield  {author} {\bibinfo {author} {\bibfnamefont {M.}~\bibnamefont
  {Eigen}}\ and\ \bibinfo {author} {\bibfnamefont {P.}~\bibnamefont
  {Schuster}},\ }\href@noop {} {\emph {\bibinfo {title} {The Hypercycle}}} ({\bibfield  {journal} {\bibinfo  {journal} {Springer}}, \bibinfo {year} {1979})}\BibitemShut {NoStop}%
\bibitem [{\citenamefont {Farmer}\ \emph {et~al.}(1986)\citenamefont {Farmer},
  \citenamefont {Kauffman},\ and\ \citenamefont
  {Packard}}]{farmer1986autocatalytic}%
  \BibitemOpen
  \bibfield  {author} {\bibinfo {author} {\bibfnamefont {J.~D.}\ \bibnamefont
  {Farmer}}, \bibinfo {author} {\bibfnamefont {S.~A.}\ \bibnamefont
  {Kauffman}}, \ and\ \bibinfo {author} {\bibfnamefont {N.~H.}\ \bibnamefont
  {Packard}},\ }\href@noop {} {\bibfield  {journal} {\bibinfo  {journal}
  {Physica D: Nonlinear Phenomena}\ }\textbf {\bibinfo {volume} {22}},\
  \bibinfo {pages} {50} (\bibinfo {year} {1986})}\BibitemShut {NoStop}%
\bibitem [{\citenamefont {Boerlijst}\ and\ \citenamefont
  {Hogeweg}(1991)}]{boerlijst1991spiral}%
  \BibitemOpen
  \bibfield  {author} {\bibinfo {author} {\bibfnamefont {M.~C.}\ \bibnamefont
  {Boerlijst}}\ and\ \bibinfo {author} {\bibfnamefont {P.}~\bibnamefont
  {Hogeweg}},\ }\href@noop {} {\bibfield  {journal} {\bibinfo  {journal}
  {Physica D: Nonlinear Phenomena}\ }\textbf {\bibinfo {volume} {48}},\
  \bibinfo {pages} {17} (\bibinfo {year} {1991})}\BibitemShut {NoStop}%
\bibitem [{\citenamefont {Jain}\ and\ \citenamefont
  {Krishna}(1998)}]{jain1998autocatalytic}%
  \BibitemOpen
  \bibfield  {author} {\bibinfo {author} {\bibfnamefont {S.}~\bibnamefont
  {Jain}}\ and\ \bibinfo {author} {\bibfnamefont {S.}~\bibnamefont {Krishna}},\
  }\href@noop {} {\bibfield  {journal} {\bibinfo  {journal} {Physical Review
  Letters}\ }\textbf {\bibinfo {volume} {81}},\ \bibinfo {pages} {5684}
  (\bibinfo {year} {1998})}\BibitemShut {NoStop}%
\bibitem [{\citenamefont {Segr{\'e}}\ \emph {et~al.}(2000)\citenamefont
  {Segr{\'e}}, \citenamefont {Ben-Eli},\ and\ \citenamefont
  {Lancet}}]{segre2000compositional}%
  \BibitemOpen
  \bibfield  {author} {\bibinfo {author} {\bibfnamefont {D.}~\bibnamefont
  {Segr{\'e}}}, \bibinfo {author} {\bibfnamefont {D.}~\bibnamefont {Ben-Eli}},
  \ and\ \bibinfo {author} {\bibfnamefont {D.}~\bibnamefont {Lancet}},\
  }\href@noop {} {\bibfield  {journal} {\bibinfo  {journal} {Proceedings of the
  National Academy of Sciences}\ }\textbf {\bibinfo {volume} {97}},\ \bibinfo
  {pages} {4112} (\bibinfo {year} {2000})}\BibitemShut {NoStop}%
\bibitem [{\citenamefont {Kaneko}(2005)}]{kaneko2005recursive}%
  \BibitemOpen
  \bibfield  {author} {\bibinfo {author} {\bibfnamefont {K.}~\bibnamefont
  {Kaneko}},\ }\href@noop {} {\bibfield  {journal} {\bibinfo  {journal}
  {Advanced in Chemical Physics}\ }\textbf {\bibinfo {volume} {130}},\ \bibinfo
  {pages} {543} (\bibinfo {year} {2005})}\BibitemShut {NoStop}%
\bibitem [{\citenamefont {Giri}\ and\ \citenamefont
  {Jain}(2012)}]{giri2012origin}%
  \BibitemOpen
  \bibfield  {author} {\bibinfo {author} {\bibfnamefont {V.}~\bibnamefont
  {Giri}}\ and\ \bibinfo {author} {\bibfnamefont {S.}~\bibnamefont {Jain}},\
  }\href@noop {} {\bibfield  {journal} {\bibinfo  {journal} {PloS One}\
  }\textbf {\bibinfo {volume} {7}},\ \bibinfo {pages} {e29546} (\bibinfo {year}
  {2012})}\BibitemShut {NoStop}%
\bibitem [{\citenamefont {Wu}\ and\ \citenamefont
  {Higgs}(2009)}]{wu2009origin}%
  \BibitemOpen
  \bibfield  {author} {\bibinfo {author} {\bibfnamefont {M.}~\bibnamefont
  {Wu}}\ and\ \bibinfo {author} {\bibfnamefont {P.~G.}\ \bibnamefont {Higgs}},\
  }\href@noop {} {\bibfield  {journal} {\bibinfo  {journal} {Journal of
  molecular evolution}\ }\textbf {\bibinfo {volume} {69}},\ \bibinfo {pages}
  {541} (\bibinfo {year} {2009})}\BibitemShut {NoStop}%
\bibitem [{\citenamefont {Togashi}\ and\ \citenamefont
  {Kaneko}(2001)}]{togashi2001transitions}%
  \BibitemOpen
  \bibfield  {author} {\bibinfo {author} {\bibfnamefont {Y.}~\bibnamefont
  {Togashi}}\ and\ \bibinfo {author} {\bibfnamefont {K.}~\bibnamefont
  {Kaneko}},\ }\href@noop {} {\bibfield  {journal} {\bibinfo  {journal}
  {Physical Review Letters}\ }\textbf {\bibinfo {volume} {86}},\ \bibinfo
  {pages} {2459} (\bibinfo {year} {2001})}\BibitemShut {NoStop}%
  \bibitem [{\citenamefont {Awazu}\ and\ \citenamefont
  {Kaneko}(2007)}]{awazu2007discreteness}%
  \BibitemOpen
  \bibfield  {author} {\bibinfo {author} {\bibfnamefont {A.}~\bibnamefont
  {Awazu}}\ and\ \bibinfo {author} {\bibfnamefont {K.}~\bibnamefont {Kaneko}},\
  }\href@noop {} {\bibfield  {journal} {\bibinfo  {journal} {Physical Review
  E}\ }\textbf {\bibinfo {volume} {76}},\ \bibinfo {pages} {041915} (\bibinfo
  {year} {2007})}\BibitemShut {NoStop}%
  \bibitem [{\citenamefont {Ohkubo}\ \emph {et~al.}(2008)\citenamefont {Ohkubo},
  \citenamefont {Shnerb},\ and\ \citenamefont
  {A.~Kessler}}]{ohkubo2008transition}%
  \BibitemOpen
  \bibfield  {author} {\bibinfo {author} {\bibfnamefont {J.}~\bibnamefont
  {Ohkubo}}, \bibinfo {author} {\bibfnamefont {N.}~\bibnamefont {Shnerb}}, \
  and\ \bibinfo {author} {\bibfnamefont {D.}~\bibnamefont {A.~Kessler}},\
  }\href@noop {} {\bibfield  {journal} {\bibinfo  {journal} {Journal of the
  Physical Society of Japan}\ }\textbf {\bibinfo {volume} {77}},\ \bibinfo
  {pages} {044002} (\bibinfo {year} {2008})}\BibitemShut {NoStop}%
\bibitem [{\citenamefont {Dauxois}\ \emph {et~al.}(2009)\citenamefont
  {Dauxois}, \citenamefont {Di~Patti}, \citenamefont {Fanelli},\ and\
  \citenamefont {McKane}}]{dauxois2009enhanced}%
  \BibitemOpen
  \bibfield  {author} {\bibinfo {author} {\bibfnamefont {T.}~\bibnamefont
  {Dauxois}}, \bibinfo {author} {\bibfnamefont {F.}~\bibnamefont {Di~Patti}},
  \bibinfo {author} {\bibfnamefont {D.}~\bibnamefont {Fanelli}}, \ and\
  \bibinfo {author} {\bibfnamefont {A.~J.}\ \bibnamefont {McKane}},\
  }\href@noop {} {\bibfield  {journal} {\bibinfo  {journal} {Physical Review
  E}\ }\textbf {\bibinfo {volume} {79}},\ \bibinfo {pages} {036112} (\bibinfo
  {year} {2009})}\BibitemShut {NoStop}%
\bibitem [{\citenamefont {Biancalani}\ \emph {et~al.}(2014)\citenamefont
  {Biancalani}, \citenamefont {Dyson},\ and\ \citenamefont
  {McKane}}]{biancalani2014noise}%
  \BibitemOpen
  \bibfield  {author} {\bibinfo {author} {\bibfnamefont {T.}~\bibnamefont
  {Biancalani}}, \bibinfo {author} {\bibfnamefont {L.}~\bibnamefont {Dyson}}, \
  and\ \bibinfo {author} {\bibfnamefont {A.~J.}\ \bibnamefont {McKane}},\
  }\href@noop {} {\bibfield  {journal} {\bibinfo  {journal} {Physical Review
  Letters}\ }\textbf {\bibinfo {volume} {112}},\ \bibinfo {pages} {038101}
  (\bibinfo {year} {2014})}\BibitemShut {NoStop}%
\bibitem [{\citenamefont {Jafarpour}\ \emph {et~al.}(2015)\citenamefont
  {Jafarpour}, \citenamefont {Biancalani},\ and\ \citenamefont
  {Goldenfeld}}]{jafarpour2015noise}%
  \BibitemOpen
  \bibfield  {author} {\bibinfo {author} {\bibfnamefont {F.}~\bibnamefont
  {Jafarpour}}, \bibinfo {author} {\bibfnamefont {T.}~\bibnamefont
  {Biancalani}}, \ and\ \bibinfo {author} {\bibfnamefont {N.}~\bibnamefont
  {Goldenfeld}},\ }\href@noop {} {\bibfield  {journal} {\bibinfo  {journal}
  {Physical Review Letters}\ }\textbf {\bibinfo {volume} {115}},\ \bibinfo
  {pages} {158101} (\bibinfo {year} {2015})}\BibitemShut {NoStop}%
\bibitem [{\citenamefont {Saito}\ and\ \citenamefont
  {Kaneko}(2015)}]{saito2015theoretical}%
  \BibitemOpen
  \bibfield  {author} {\bibinfo {author} {\bibfnamefont {N.}~\bibnamefont
  {Saito}}\ and\ \bibinfo {author} {\bibfnamefont {K.}~\bibnamefont {Kaneko}},\
  }\href@noop {} {\bibfield  {journal} {\bibinfo  {journal} {Physical Review
  E}\ }\textbf {\bibinfo {volume} {91}},\ \bibinfo {pages} {022707} (\bibinfo
  {year} {2015})}\BibitemShut {NoStop}%
  \bibitem [{abo({\natexlab{a}})}]{about-unit}%
  \BibitemOpen
 Unit of polymer concentration is $\mathrm{[number/L]}$. 
 If we use $\mathrm{[mol/L]}$ 
 , we need to make a unit conversion by multiplying by the Avogadro constant.
  \bibitem [{\citenamefont {Van~Kampen}(1992)}]{van1992stochastic}%
  \BibitemOpen
  \bibfield  {author} {\bibinfo {author} {\bibfnamefont {N.~G.}\ \bibnamefont
  {Van~Kampen}},\ }\href@noop {} {\emph {\bibinfo {title} {Stochastic processes
  in physics and chemistry}}},\ Vol.~\bibinfo {volume} {1}\ (\bibinfo
  {publisher} {Elsevier},\ \bibinfo {year} {1992})\BibitemShut {NoStop}%
\bibitem [{\citenamefont {Gillespie}(1977)}]{gillespie1977exact}%
  \BibitemOpen
  \bibfield  {author} {\bibinfo {author} {\bibfnamefont {D.~T.}\ \bibnamefont
  {Gillespie}},\ }\href@noop {} {\bibfield  {journal} {\bibinfo  {journal} {The
  journal of physical chemistry}\ }\textbf {\bibinfo {volume} {81}},\ \bibinfo
  {pages} {2340} (\bibinfo {year} {1977})}\BibitemShut {NoStop}%
 \bibitem [{abo({\natexlab{b}})}]{about-unitconversion}%
  \BibitemOpen
With the unit of  concentration [mol/L], this expressions is written as $V_{opt} \sim 1/(x^b_4 N_A)$ \cite{about-unit}.
  \bibitem [{\citenamefont {Gardiner}\ \emph {et~al.}(1985)\citenamefont
  {Gardiner} \emph {et~al.}}]{gardiner1985handbook}%
  \BibitemOpen
  \bibfield  {author} {\bibinfo {author} {\bibfnamefont {C.~W.}\ \bibnamefont
  {Gardiner}} \emph {et~al.},\ }\href@noop {} {\emph {\bibinfo {title}
  {Handbook of stochastic methods}}},\ Vol.~\bibinfo {volume} {4}\ (\bibinfo
  {publisher} {Springer Berlin},\ \bibinfo {year} {1985})\BibitemShut {NoStop}%
\bibitem [{\citenamefont {Gillespie}(2000)}]{gillespie2000chemical}%
  \BibitemOpen
  \bibfield  {author} {\bibinfo {author} {\bibfnamefont {D.~T.}\ \bibnamefont
  {Gillespie}},\ }\href@noop {} {\bibfield  {journal} {\bibinfo  {journal} {The
  Journal of Chemical Physics}\ }\textbf {\bibinfo {volume} {113}},\ \bibinfo
  {pages} {297} (\bibinfo {year} {2000})}\BibitemShut {NoStop}%
\bibitem [{\citenamefont {Duncan}\ \emph {et~al.}(2015)\citenamefont {Duncan},
  \citenamefont {Liao}, \citenamefont {Vejchodsk{\`y}}, \citenamefont {Erban},\
  and\ \citenamefont {Grima}}]{duncan2015noise}%
  \BibitemOpen
  \bibfield  {author} {\bibinfo {author} {\bibfnamefont {A.}~\bibnamefont
  {Duncan}}, \bibinfo {author} {\bibfnamefont {S.}~\bibnamefont {Liao}},
  \bibinfo {author} {\bibfnamefont {T.}~\bibnamefont {Vejchodsk{\`y}}},
  \bibinfo {author} {\bibfnamefont {R.}~\bibnamefont {Erban}}, \ and\ \bibinfo
  {author} {\bibfnamefont {R.}~\bibnamefont {Grima}},\ }\href@noop {}
  {\bibfield  {journal} {\bibinfo  {journal} {Physical Review E}\ }\textbf
  {\bibinfo {volume} {91}},\ \bibinfo {pages} {042111} (\bibinfo {year}
  {2015})}\BibitemShut {NoStop}%
\bibitem [{\citenamefont {Haruna}(2010)}]{haruna2010investigating}%
  \BibitemOpen
  \bibfield  {author} {\bibinfo {author} {\bibfnamefont {T.}~\bibnamefont
  {Haruna}},\ }\href@noop {} {\bibfield  {journal} {\bibinfo  {journal}
  {Journal of Computer Chemistry, Japan}\ }\textbf {\bibinfo {volume} {9}},\
  \bibinfo {pages} {135} (\bibinfo {year} {2010})}\BibitemShut {NoStop}%
\bibitem [{\citenamefont {Luisi}(2006)}]{luisi2006emergence}%
  \BibitemOpen
  \bibfield  {author} {\bibinfo {author} {\bibfnamefont {P.~L.}\ \bibnamefont
  {Luisi}},\ }\href@noop {} {\emph {\bibinfo {title} {The emergence of life:
  from chemical origins to synthetic biology}}}\ (\bibinfo  {publisher}
  {Cambridge University Press},\ \bibinfo {year} {2006})\BibitemShut {NoStop}%
\bibitem [{\citenamefont {Rasmussen et.al.}(2009)}]{rasmussen2009protocells}%
  \BibitemOpen
  \bibfield  {author} {\bibinfo {author} {\bibfnamefont {S.}~\bibnamefont
  {Rasmussen} \emph {et~al.} },\ }\href@noop {} {\emph {\bibinfo {title} {Protocells}}}\
  (\bibinfo  {publisher} {Mit Press},\ \bibinfo {year} {2009})\BibitemShut
  {NoStop}%
\bibitem [{\citenamefont {Graham}\ and\ \citenamefont
  {T{\'e}l}(1984)}]{graham1984weak}%
  \BibitemOpen
  \bibfield  {author} {\bibinfo {author} {\bibfnamefont {R.}~\bibnamefont
  {Graham}}\ and\ \bibinfo {author} {\bibfnamefont {T.}~\bibnamefont
  {T{\'e}l}},\ }\href@noop {} {\bibfield  {journal} {\bibinfo  {journal}
  {Journal of statistical physics}\ }\textbf {\bibinfo {volume} {35}},\
  \bibinfo {pages} {729} (\bibinfo {year} {1984})}\BibitemShut {NoStop}%
\end{thebibliography}

%

\end{document}